# Discovery of magnetic field line dependent anisotropic chemiresistive response in Magnetite: A new piece to the puzzle of magnetoreception

**Pratyasha Rudra[1,2] and Swastik Mondal[1,2*]**

[1]CSIR-Central Glass and Ceramic Research Institute; Jadavpur, Kolkata-700032, India

[2]Academy of Scientific and Innovative Research (AcSIR); Ghaziabad- 201002, India

*E-mail: swastik_mondal@cgcri.res.in

## Abstract

Chemiresistive materials, which alter their electrical resistance in response to interactions with surrounding chemicals, are valued for their robustness, rapid detection ability and high sensitivity. Recent research has revealed that the sensing performance of these materials can be enhanced by applying an external magnetic field. In this study, we report a novel finding in the chemiresistive behavior of magnetite ($Fe_3O_4$), where its response has been found to be modulated in an anisotropic manner when exposed to an external magnetic field, analogous to Earth's magnetic field. Remarkably, substantial variations have been observed in response to analytes naturally present in the atmosphere. This chemiresistive response exhibits a strong anisotropic dependence on the strength, direction and inclination of the magnetic field, suggesting that magnetite's electrical resistance dynamically adapts to both magnetic and chemical environmental changes. The observed behavior under Earth-like magnetic field closely mirrors the magnetoreception seen in migratory species that rely on magnetite for navigation. This finding may provide new insights into the mechanisms behind magnetite-based magnetoreception observed in various biological species.

## Introduction

Magnetite, a naturally occurring iron oxide ($Fe_3O_4$), possesses remarkable magnetic properties, making it a subject of great interest in the field of magnetoreception [1-5]. Within biological species, including bacteria, fish, birds, mammals and certain invertebrates, Magnetite particles have been identified and observed to form crystalline structures [5-10]. Several theories have been put forth in an attempt to explain how Magnetite contributes to magnetoreception, including the single-domain Magnetite alignment theory, magnetic particle motion theory, Magnetite chain hypothesis and biocompass hypothesis [9, 11-15]. Each of these theories provides a distinct perspective on how Magnetite may influence magnetoreception, suggesting mechanisms ranging from particle alignment in response to Earth's magnetic field to considerations of their motion and electronic properties. Surprisingly, while several properties of Magnetite have been extensively studied, its magneto-chemiresistance — how its electrical resistance changes upon exposure to analytes under the influence of magnetic field — has remained an unexplored territory. Given that Magnetite particles are present in the olfactory tissues of Fish [6, 16, 17], in the upper beaks of birds [4, 18, 19], in the antennae and abdomen of insects [20, 21] and so on, these particles are therefore likely to be exposed to external chemicals present in the environment. Therefore, it is imperative to explore the chemiresistive response of Magnetite in the presence of analytes found in the environment under the influence of a magnetic field. Such studies could shed light on how Magnetite particles interact with their chemical surroundings and provide valuable insights into the complex interplay between Magnetite, magnetic fields and chemical cues in the context of magnetoreception.

Recently, it was discovered that application of magnetic field can alter the sensing response of chemiresistive materials [22, 23] — a type of sensing materials whose electrical resistance varies with changes in the surrounding chemical environment. This discovery of magneto-chemiresistive property has been leveraged to enhance the chemical detection capabilities of various materials, particularly of metal oxide semiconductors with residual magnetic moments [22-25]. Given that $Fe_3O_4$ is a metal oxide semiconductor with intrinsic ferrimagnetism, it is expected to exhibit a significant level of magneto-chemiresistive sensing response. In this study, we systematically examined variations in Magnetite's chemiresistive sensing response when exposed to various analytes present in the atmosphere, both in the presence and absence of external magnetic fields. Additionally, we rigorously assessed how Magnetite's chemiresistive response is influenced by variations in magnetic field strength and orientation of magnetic field lines.

## Characterizations of Magnetite particles and development of a set-up for controlled magnetic environment

To investigate the role of strength, direction and inclination of magnetic field lines analogous to that of the Earth's on the sensing response of Magnetite, the sample was prepared through a hydrothermal route (Supplementary Methods) and the formation of Magnetite was confirmed by X-ray diffraction (XRD) experiment (Fig. 1A). Afterwards, transmission electron microscopy (TEM), field emission scanning electron microscopy (FESEM)-energy dispersive X-ray (EDX) and X-ray photoelectron spectroscopy (XPS) studies were conducted to examine the morphology of the Magnetite particles and the presence of each distinct element including their oxidation states (Fig. 1, B to I and fig. S1). Magnetic measurements were performed to confirm the ferrimagnetic nature of the Magnetite sample (Fig. 1J). Prior to chemiresistive sensing experiments, Hall measurements were performed to identify the sample as an n-type semiconductor (see supplementary text). Subsequently, the Magnetite sample was deposited onto a flat alumina substrate featuring interdigitated gold electrodes by drop casting method to facilitate chemiresistive sensing measurements (Fig. 2, A and B). To ensure that all experiments were conducted within a controlled magnetic environment, free from external magnetic influences, including the Earth's magnetic field, a custom chamber was constructed using mu-metal (designated as the mu-chamber, Fig. 2C). Mu-metal, an alloy composed primarily of nickel and iron, exhibits high magnetic permeability, effectively blocking the passage of magnetic field lines [28]. The magnetic field within the mu-chamber was verified to be effectively zero using a Gauss meter. As the mu-chamber is fully enclosed on all sides (Fig. 2C), its interior is entirely dark. To simulate controlled magnetic field environment, a bar magnet was employed. The bar magnet and the Magnetite sensor were positioned on a wooden base at a specified distance from one another (Fig. 2D). The strength of the magnetic field at the Magnetite sample could be adjusted by varying the distance between the Magnetite sensor and the bar magnet. All chemiresistive sensing experiments were conducted at room temperature under controlled flow of analytes within this mu-chamber (Fig. 2, D to G).

## Chemiresistive behavior of Magnetite in absence of a magnetic field

For selection of analytes, the obvious choices were atmospheric air and its major components — nitrogen ($N_2$), oxygen ($O_2$), carbon dioxide ($CO_2$) and water vapor. In addition to the primary natural atmospheric constituents, nitrogen dioxide ($NO_2$) was chosen as an analyte to serve as a representative of key pollutants present in the modern Earth's atmosphere. Under zero magnetic field strength, purging with air resulted in a noticeable decrease in chemiresistance of the Magnetite sample (fig. S2A). The responses to pure oxygen and nitrogen were negligible, while for $CO_2$ and $NO_2$, slight responses have been observed (fig. S2, B to D, table S1). In addition to almost inert nitrogen, air predominantly contains oxygen, which is an oxidizing agent expected to increase the resistance of n-type Magnetite. However, the decrease in resistance observed (fig. S2A) for purging air suggests that Magnetite may be responsive towards humidity. To explore this, we measured the chemiresistive responses to humid air with 98% relative humidity (RH) and dry air with 7% RH, finding slight responses of 0.58(2)% and 0.48(2)% respectively under zero magnetic field (fig. S2, E to F and table S1). Among all analytes tested, magnetite has shown highest response to 20 ppm $NO_2$ under zero magnetic field (table S1).

## Influence of magnetic field strength on chemiresistive performance of Magnetite

In order to investigate the effect of magnetic field strength on chemiresistive sensing response, a controlled magnetic field environment was created around the Magnetite sensor inside the mu-chamber using a bar magnet (Fig. 2D). The bar magnet was oriented parallel to Earth's surface with its north pole aligned towards the geomagnetic north (Fig. 2F). The distance of the bar magnet from the sample was adjusted to ensure that the magnetic field strength at the Magnetite sample was 0.05(1) mT (Fig. 2D), which approximates the average geomagnetic field strength [27, 28]. Since water vapor is the third most abundant component in air (up to ~4%), following $N_2$ and $O_2$, we started our magnetic field-induced sensing experiments by focusing on RH. This arrangement resulted in a significant response value of 15.33(26)% towards air with 98% RH, a ~26-fold improvement in magneto-chemiresistive response compared to the response without any magnetic field. To further explore the impact of magnetic field strength on the chemiresistive response of Magnetite to 98% RH, additional experiments were conducted with different field strengths by varying the distance between the bar magnet and the sample, while keeping all other experimental parameters constant. The responses to 98% RH exhibited variation depending on the strength of the applied magnetic field (Fig. 3A and table S2). These findings suggest that the magneto-chemiresistive response of Magnetite increases up to a certain magnetic field strength, beyond which it begins to decrease. A similar trend has been recently observed in the magneto-chemiresistive $NO_2$ sensing response of Fe-doped indium oxide [25].

## Direction of field lines can drastically alter chemiresistive response of Magnetite

Under Earth like magnetic field, decrease in response observed when the magnetic field line direction changed from geomagnetic north to geomagnetic northeast. By rotating the magnet clockwise in the same plane (Fig. 2F), the angle between the direction of charge flow in the sample and the magnetic field lines changed, leading to anisotropic magneto-chemiresistive response. Further decreases occurred when the field line direction shifted to the geomagnetic east. The response increased when the field lines aligned toward geomagnetic southeast, peaking at 12.51(25)% when aligning parallel to geomagnetic south. This trend was consistent for each quadrant of the geomagnetic axes. Although the responses remained nearly similar for opposite geomagnetic directions (table S3), the change in response of Magnetite to 98% RH with variations in field line direction is clearly evident (Fig. 3B and fig. S3). These results may provide a mechanistic basis by which Magnetite particles may be used to identify geomagnetic direction. Similar change in magneto-chemiresistive response of Magnetite towards $NO_2$ and 7% RH have been observed (Fig. 3C and fig. S4 and table S4), however, the absolute values of sensing responses were different. The observation that the chemiresistive response varies with changes in RH is noteworthy, given that humidity is a well-documented factor influencing and initiating migration in various biological species [29, 30].

## Assessing how the sensing response of Magnetite is affected by the inclination of magnetic field lines

It is known that biological species that use magnetoreception can recognize the inclination of the magnetic field lines [2, 31-34]. In order to investigate the effect of magnetic field line inclination on the magneto-chemiresistive response of Magnetite, experiments were conducted by positioning the Magnetite-containing flat substrate at various angles relative to the base of the mu-chamber (Fig. 2G). The position of the bar magnet was maintained parallel to the base throughout the experiment with its north pole aligned towards the geomagnetic north generating 0.05(1) mT field strength on the Magnetite sample. By varying the substrate's angle from 0° to 90° relative to the base, different inclination angles of the magnetic field lines were effectively created on the Magnetite sample (Fig. 2G). The magneto-chemiresistive response to humid air (98% RH) have been measured at angles 0°, 22.5°, 45°, 67.5° and 90°. The highest response of 15.33(26)% was observed when the sensor is positioned at 0°, parallel to the bar magnet in the same horizontal plane. The response drastically decreases with increasing inclination angle with lowest response of 1.86(22)% observed for 90° (Fig. 3D and fig. S5 and table S3). Similarly, for 20 ppm $NO_2$, the response of magnetite significantly decreased as the inclination angle of the magnetite sensor increased (Fig. 3E). The highest response of 3.54(23), was observed when the sensor was positioned at 0°, parallel to the bar magnet in the same horizontal plane. The response greatly declined to 1.13(8) when the sensor was inclined at 90°. Thus, in addition to the strength of the magnetic field and the directions of the magnetic field lines, the magneto-chemiresistive response of Magnetite is also greatly influenced by the inclination angle of the magnetic field lines. These observations thus provide telltale insights into how Magnetite-based magnetoreception in biological organisms may interpret both the direction and inclination of magnetic field lines [33, 34].

**Influence of light on the sensing response of Magnetite**

All previous experiments were conducted in the dark, within the mu-chamber. In the final set of experiments, controlled illumination using LED sources was introduced inside the mu-chamber. Exposure of white light (Supplementary Methods) resulted in a substantial decrease in the magneto-chemiresistive 98% RH sensing response (table S3), though the overall pattern remains same (Fig. 3F and fig. S6 and table S3). Notable variations in the magneto-chemiresistive response of Magnetite were observed with lights of different wavelengths (fig. S7 and table S5). Variation in response has also been observed in the case of 20 ppm $NO_2$ sensing by magnetite when exposed to lights of different wavelengths (Fig. S8). The modification of chemiresistive sensing responses due to light excitation is well-documented in the literature [35]. This phenomenon is primarily attributed to changes in electron mobility [36], alterations in surface reaction dynamics [37] and modifications in the bandgap [38] of the chemiresistive material resulting from photoexcitation. Therefore, these findings suggest that the chemiresistive response of Magnetite will be influenced by both magnetic fields and light. This is particularly noteworthy as it has been reported that modifying magnetic field and light wavelength do affect magnetoreception in certain biological species [39-42].

**Discussion**

The magneto-chemiresistive response of Magnetite to $CO_2$ exhibits similar patterns to those observed towards RH and $NO_2$; however, the absolute values of the sensing responses differ (fig. S9A). Figure S9B shows that magnetite is basically selective towards $NO_2$ gas, probably because of the paramagnetic nature of $NO_2$. The presence of unpaired spins allows more $NO_2$ molecules to engage in charge exchange with Magnetite particles when subjected to an attractive magnetic field, compared to the other analytes tested in this study. This likely contributes to the highest response observed for 20 ppm $NO_2$ under the magnetic field. The observations of similar patterns of magneto-chemiresistive response of Magnetite towards various analytes (Fig. 3, B and C and fig. S9A) suggest that, within a given chemical environment, Magnetite will display anisotropic magneto-chemiresistive responses irrespective of the specific analytes involved. As illustrated in Fig. 3, B and C, the plot of the angle-dependent anisotropic magneto-chemiresistance response closely resembles a typical anisotropic magnetoresistance plot [43, 44]. This similarity arises because the magneto-chemiresistive response measures the variance in electrical resistance in the presence and absence of analytes under a magnetic field (fig. S10), which can be directly correlated to magnetoresistance (see supplementary text, tables S6 and S7). Magnetoresistance is a well-established physical phenomenon in which the electrical resistance of a material varies in response to an applied magnetic field [43, 44]. This effect is utilized in various technologies, including magnetic sensors [44]. The principal distinction between magnetoresistance and magneto-chemiresistance is that the latter involves an additional charge exchange between the sensing material and surrounding chemicals. Magnetoresistance primarily consists of two components: Ordinary Magnetoresistance (OMR) and Anisotropic Magnetoresistance (AMR) [45]. OMR arises predominantly from the Lorentz force and the cyclotron motion of charge carriers within a magnetic field, resulting in an isotropic change in resistance [45]. In contrast, AMR induces an anisotropic variation in resistance, which depends on the orientation of the magnetic field lines relative to the material. AMR is fundamentally a quantum mechanical effect, driven by spin-dependent scattering in magnetic materials [46, 47]. Since AMR governs the directional changes in electrical resistance in magnetic materials, it is expected that magneto-chemiresistive sensing responses or any other electrical signals involving Magnetite particles, will be influenced by the orientation of the Earth's magnetic field.

The discovery of anisotropic magneto-chemiresistive property in Magnetite offers a crucial answer to the pressing question of whether Magnetite has the essential physical and magnetic properties to operate effectively as a magnetosensor in biological systems [4]. Behavioral studies of species known to utilize Magnetite for magnetoreception have revealed various unresolved issues [3, 15, 48]. Notably, there is currently no satisfactory explanation for how Magnetite perceives the inclination of geomagnetic field lines [3]. While it is established that species containing biogenic Magnetite are unable to differentiate between geomagnetic north and south at the geomagnetic equator, they can distinguish between the poles in the Northern and Southern Hemispheres [3, 48]. The present study demonstrates that similar magneto-chemiresistive responses to RH are observed for opposite geomagnetic directions (Fig. 3B and fig. S3 to S4 and table S3 to S4). This suggests that distinguishing between geomagnetic north and south — or geomagnetic east and west — becomes challenging when the magnetic field lines are aligned parallel to the Magnetite sensor (i.e., at an inclination angle of 0°), as occurs at the geomagnetic equator. In contrast, in the Northern and Southern Hemispheres, where the inclination angle of the magnetic field lines varies, it is feasible to differentiate between opposing geomagnetic directions.

Another unresolved issue pertains to why migratory species that possess biogenic Magnetite, such as the European Robin, become disoriented by sudden exposure to light during migration [41]. In contrast, these species maintain their orientation when exposed to the same light prior to migration [40]. The present findings indicate (Fig. 3F and fig. S7 and table S3 and S5) that exposure to light reduces the absolute values of magneto-chemiresistive sensing response of Magnetite; however, the direction-dependent response pattern remains overall consistent (Fig. 3B and F and fig. S7 and table S3 and S5). This suggests that sudden exposure to light reduces the sensor's response, potentially leading to confusion in the migratory species regarding the orientation of magnetic field lines. Conversely, prior exposure to light allows the species to acclimate to the diminished responses, thus preventing disorientation during migration.

These findings thus indicate that species employing Magnetite-based magnetoreception likely integrate both magnetic and chemical cues for navigation. It is expected that, other processes that involve interactions with external analytes, such as electrochemical sensing, will also exhibit similar effects under magnetic fields. This mechanism proposes an intriguing analogy: it suggests that these organisms may "smell" magnetic field lines, revealing a sophisticated integration of sensory modalities in their migratory behavior. The correlation between behavioral patterns of species that have biogenic Magnetite and the anisotropic magneto-chemiresistive response of Magnetite presents a compelling case that this physical mechanism may help bridge existing gaps in our understanding of Magnetite-based magnetoreception.

**Acknowledgement**

P.R. expresses gratitude to DST for awarding the INSPIRE fellowship (DST/INSPIRE Fellowship/2018/IF180761). S.M. acknowledges financial support from the SERB Core Research Grant, Government of India (Grant numbers: CRG/2023/002697). Authors thank Materials Characterization & Instrumentation Division for XPS, FESEM-EDX and TEM facility. Authors would like to thank Pratap Das and the members of Engineering Services Division of CSIR-CGCRI for technical help. Authors would like to thank Centre for Materials Characterization unit and Physical and Materials Chemistry Division of NCL, Pune for XPS and SQUID facility. Authors acknowledge Electrical Characterization Facility, IIT-Kanpur for Hall measurement facility.

**Contributions**

P.R.: methodology; software; validation; formal analysis; investigation; writing – original draft; writing – review and editing; visualization. S.M.: conceptualization; methodology; validation; formal analysis; data curation; writing – original draft; writing – review and editing; visualization; supervision; resources; funding acquisition.

**Competing interests:** Authors declare that they have no competing interests.

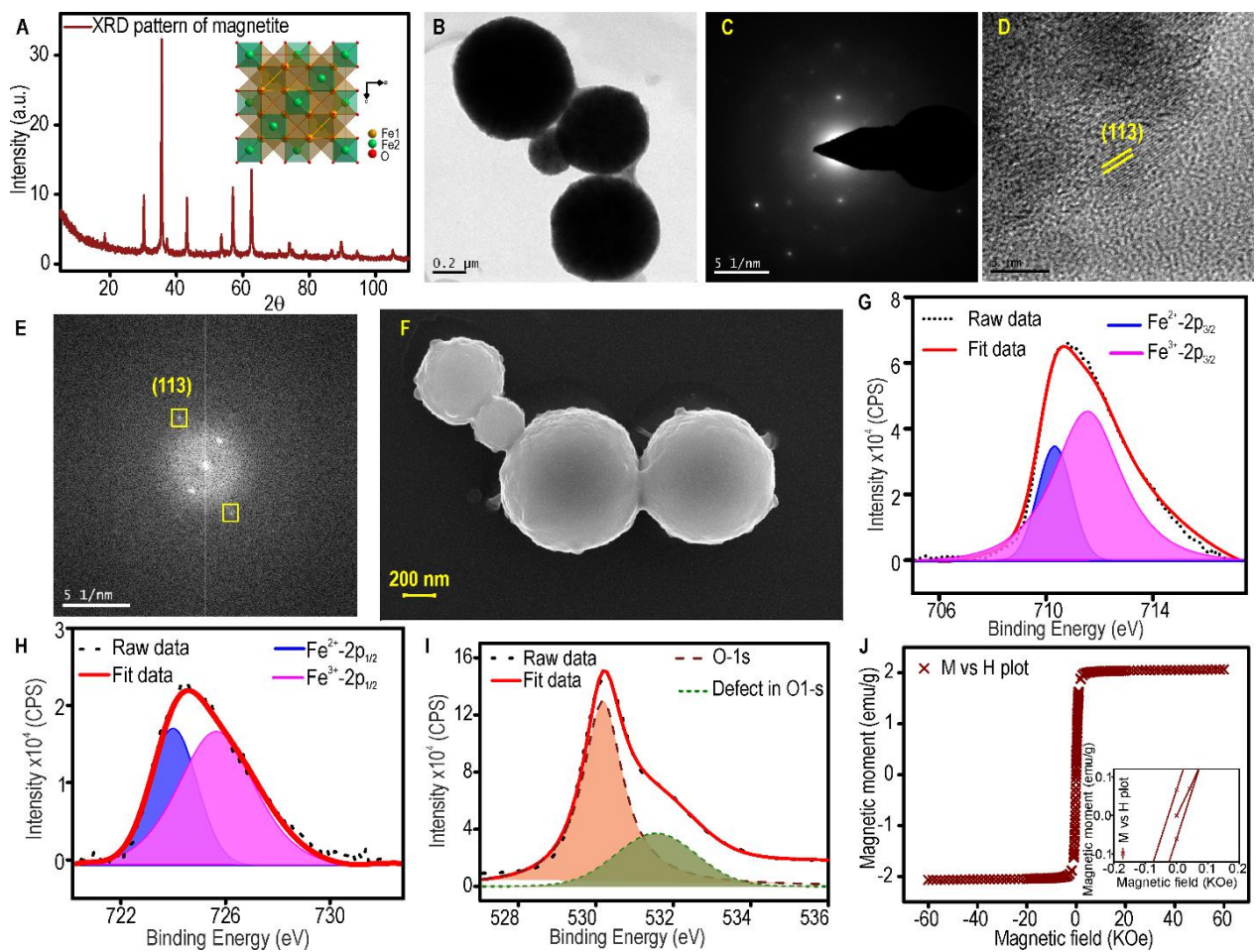

**Fig. 1. Characterization results.** (**A**) XRD pattern of Magnetite and its crystal structure (inset). (**B**) TEM image of nano-spheres in Magnetite. (**C**) SAED pattern from Magnetite nano-spheres. (**D** and **E**) (113) planes of crystalline Magnetite observed using HRTEM and corresponding FFT. (**F**) FESEM image of Magnetite nano-spheres. (**G**) Fe-2p$_{3/2}$ spectra from XPS, which have been deconvoluted in two distinct peaks corresponding to Fe$^{2+}$ highlighted in blue and Fe$^{3+}$ highlighted in magenta. (**H**) Fe-2p$_{1/2}$ spectra from XPS, that have been deconvoluted in two distinct peaks corresponding to Fe$^{2+}$ highlighted in blue and Fe$^{3+}$ highlighted in magenta. (**I**) O-1s spectra from XPS where O1s lattice spectra has been shown in peach and vacancy in O1s lattice has been indicated in olive green color. (**J**) M-H plot of Magnetite.

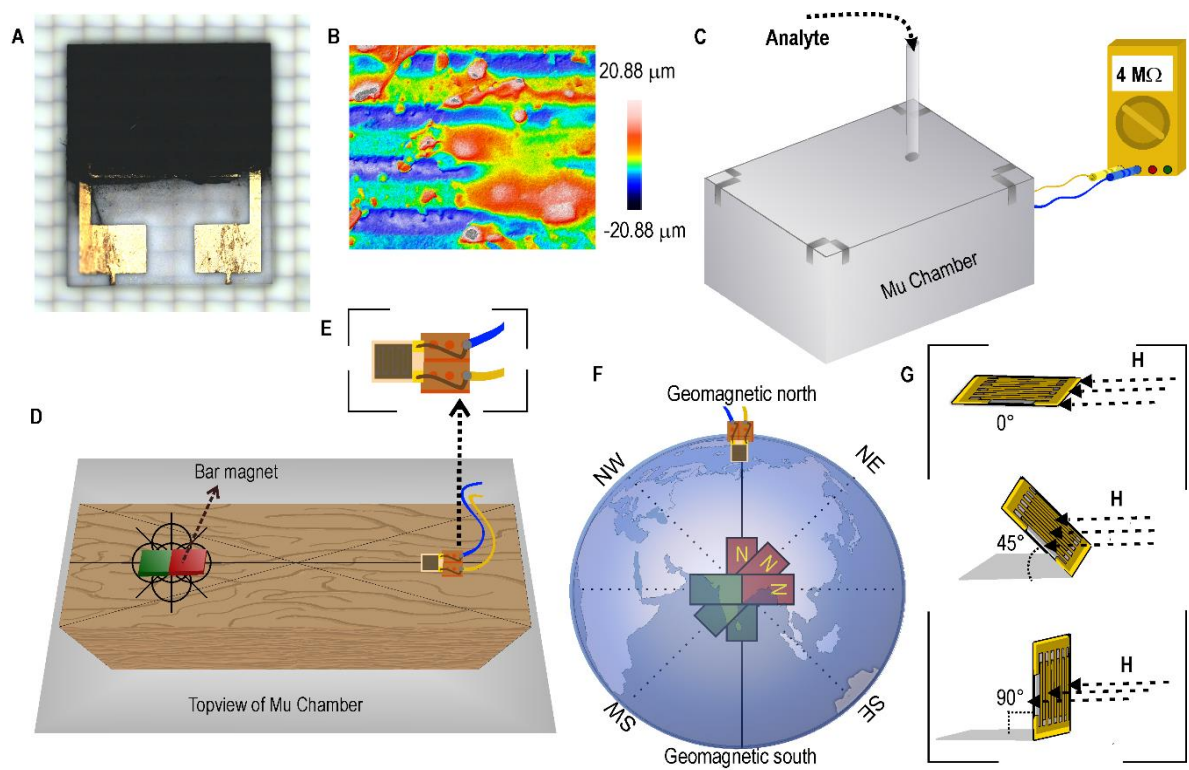

**Fig. 2. Magneto-chemiresistive sensing experiment set up using Magnetite sensor.** (**A**) Magnetite coated on alumina substrate with interdigitated gold electrodes. (**B**) Surface topology of Magnetite sensor. (**C**) Mu chamber used for magnetic field-controlled gas sensing experiment with a gas inlet on top, wherein the Magnetite sensor is connected to a multimeter. (**D**) Arrangement of the Magnetite sensor and the bar magnet inside the Mu chamber for magneto-chemiresistive sensing experiments: Magnetite sensor and the bar magnet are placed on a wooden block at a distance on the Mu chamber base. (**E**) Enlarged image of Magnetite sensor connected to a flat stripboard. (**F**) Schematic illustration of field line direction dependent magneto-chemiresistive sensing experiment: direction of bar magnet has been changed clockwise along geomagnetic cardinal and ordinal directions to change direction of field line passing through the Magnetite sensor. (**G**) Schematic illustration of orientation of the sensor during inclination angle dependent magneto-chemiresistive sensing experiment.

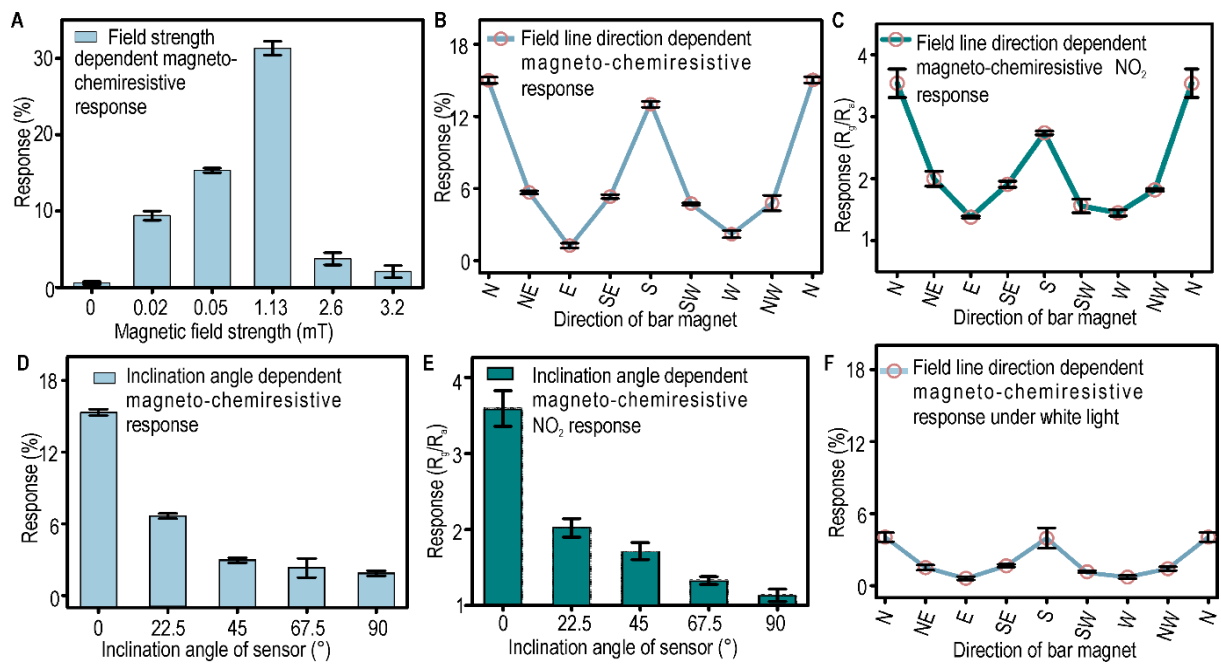

**Fig. 3. Magneto-chemiresistive response of Magnetite to 98% RH and 20 ppm NO$_2$**. (**A**) Field strength dependent magneto-chemiresistive sensing response of Magnetite. (**B**) Field line direction dependent magneto-chemiresistive sensing response of Magnetite to 98% RH. (**C**) Field line direction dependent magneto-chemiresistive sensing response of Magnetite to 20 ppm NO$_2$. (**D**) Inclination angle dependent magneto-chemiresistive sensing response of Magnetite to 98% RH. (**E**) Inclination angle dependent magneto-chemiresistive sensing response of Magnetite to 20 ppm NO$_2$. (**F**) Field line direction dependent magneto-chemiresistive sensing response of Magnetite under white light. (Directions indicated in each fig. are denoted using conventional abbreviations).

# Discovery of magnetic field line dependent anisotropic chemiresistive response in Magnetite: A new piece to the puzzle of magnetoreception

**Pratyasha Rudra[1,2] and Swastik Mondal[1,2*]**

[1]CSIR-Central Glass and Ceramic Research Institute; Jadavpur, Kolkata-700032, India

[2]Academy of Scientific and Innovative Research (AcSIR); Ghaziabad- 201002, India

*E-mail: swastik_mondal@cgcri.res.in

**The PDF file includes:**

1. Supplementary Methods
2. Supplementary Discussion
3. Supplementary Figures (Figs. S1 to S10)
4. Supplementary Tables (Tables S1 to S7)

**Supplementary Methods**

Synthesis of Magnetite Nanoparticles

Magnetite ($Fe_3O_4$) nanoparticles were synthesized using a hydrothermal method [49]. Iron (III) chloride hexahydrate ($FeCl_3 \cdot 6H_2O$) was used as the iron precursor. A solution was prepared by dissolving $FeCl_3 \cdot 6H_2O$, polyethylene glycol and sodium acetate in ethylene glycol at a molar ratio of 1:2:0.7, respectively, in 90 mL of ethylene glycol. The mixture was stirred continuously for 5 hours to ensure homogeneity. Subsequently, the solution was transferred to a Teflon-lined autoclave, which was then sealed and placed in an oven for heating at 180 °C for 18 hours. After the completion of the hydrothermal reaction, the resultant mixture was subjected to centrifugation at 10,000 rpm for 10 minutes. The precipitate was washed multiple times with an aqueous ethanol solution, followed by additional centrifugation to remove any remaining impurities. The purified product was then dried at 60 °C for 20 hours in a vacuum oven, yielding the final magnetite powder.

Sample Preparation for Characterization

Powdered samples were utilized for X-ray diffraction analysis. For microstructural studies, 2 mg of magnetite powder was dispersed in 3 mL of isopropyl alcohol. The resulting dispersion was then drop-cast onto carbon-coated copper grids and 4×4 mm² glass substrates for TEM and FESEM analysis. Additionally, three pellets were prepared from the powdered magnetite using a hydraulic press, which were subsequently used for EDX, XPS and Hall experiment.

Characterization of Magnetite Nanoparticles

The primary phase identification of the magnetite powder was conducted using an in-house X'pert Pro X-ray diffractometer (PANalytical) (Fig. 1A). Microstructural analysis was performed with a Tecnai G2 30ST (FEI) TEM (Fig. 1B). HRTEM and corresponding fast Fourier transform (FFT) techniques were employed to further observe the lattice planes (Fig. 1, D and E). Additionally, selected area electron diffraction (SAED) patterns were recorded using the same microscope (Fig. 1C). Surface morphology was examined using a Carl Zeiss Sigma 35 VP field emission scanning electron microscope (Fig. 1F). To confirm the chemical composition, EDX analysis was also conducted using the same microscope (fig. S1). Quantitative analysis of the surface electronic structure of all samples was performed using X-ray photoelectron spectroscopy (XPS) (Fig. 1 G to I). Data were acquired with a PHI 5000 VersaProbe II instrument (Physical Electronics). During the measurements, photoelectrons emitted from the sample surfaces were irradiated with aluminum Kα X-rays (photon energy:

1486.7 eV). The kinetic energy of these photoelectrons was measured at a pass energy of 11.75 eV with a step size of 0.10 eV at room temperature. The magnetic property of magnetite (Fig. 1J) was examined using a SQUID-vibrating sample magnetometer (SQUID-VSM) from Quantum Design, USA. Prior to measurements, the instrument was calibrated using a high-purity palladium metal standard at 298 K in a 1 T external magnetic field, operated in VSM mode. This calibration ensures that the measured magnetization of the standard closely matched its theoretical value, with an accuracy within 0.002 mT. The charge carrier concentration of the magnetite sample was investigated using an Ecopia HMS 5000 table top Hall measurement system (equipped with a built-in temperature control) by employing the Van der Pauw method at room temperature. A 1 cm diameter pellet, prepared from magnetite powder sample, was used for the experiment. A source current of 1 nA and a magnetic field of 0.545 T was applied during the measurement.

Fabrication and Testing of Chemiresistive Gas Sensors

Chemiresistive thick-film gas sensors were fabricated by drop-casting a slurry of freshly synthesized magnetite powder onto flat alumina substrates (3 × 5 mm$^2$) featuring gold interdigitated electrodes. An image of the magnetite sensor and its surface topography were captured and analyzed using the Sensoscan S Neox noncontact optical profilometer in focus variation mode (20x resolution) (Fig. 2A to B). The topography has been investigated on a surface area of 845.94×705.18 μm$^2$.

For gas sensing experiments, a fume hood was utilized with gases supplied from commercially available calibrated cylinders. Gas flow was precisely controlled using an Alicat mass flow controller (MFC). During the sensing experiments, dynamic resistance changes were monitored using a Keysight 34470A 7-1/2 digit multimeter interfaced with Keysight BenchVue software. Sensor responses (R) to reducing and oxidizing analytes were calculated using equation S1 and S2 respectively.

Response, R= ($R_a$-$R_g$)/$R_a$                                             (S1)

Response, R= $R_g$/$R_a$                                                    (S2)

Where, $R_a$ and $R_g$ are resistances of magnetite in air and gas respectively.

Magneto-chemiresistive Sensing Experiments in a Controlled Magnetic Field zone

A Mu-chamber with dimensions of 11 × 8 × 8 inch$^3$ (L×W×H) was constructed to conduct sensing experiments under controlled relative humidity and magnetic field conditions. The chamber was equipped with a 5 mm diameter inlet on its lid through which a gas pipe passed directly, allowing for precise purging of relative humidity and gases onto the sensor surface. Relative humidity (%RH) environments were created using saturated salt solutions, providing humidity levels of 98% and 7%. However, the actual value of RH may slightly vary at the time of sensing experiments due to the presence of ambient air inside the mu chamber. The magnetic field inside the chamber was generated using a bar magnet. The substrate was positioned at a fixed distance to experience a magnetic field strength similar to that of the Earth's. The strength of the field on the magnetite sensor was measured using a Lutron Mg-3002 AC/DC magnetometer. The magnetic field strength at the magnetite sensor was adjusted by varying the distance between the magnet and the sensor. For direction-dependent sensing experiments, the north pole of the bar magnet was rotated clockwise relative to the geomagnetic north. Additionally, LED lights were used to create an illuminated zone around the magnetite sensor during direction-dependent sensing experiments conducted under light. For light-dependent magneto-chemiresistive gas sensing, LEDs with intensity of 1800 mcd were employed. For experiments involving LEDs of higher intensity, LEDs with an intensity of 3600 mcd were used. The dominant wavelength ranges of the LEDs used were as follows: white LEDs (380-750 nm), blue LEDs (450-460 nm), green LEDs (510-530 nm), and red LEDs (630-635 nm).

**Supplementary Discussion**

Crystal and Microstructure

X-ray diffraction (XRD) data reveals that the magnetite exhibits a cubic structure with an Fd-3m (1) space group and a lattice parameter of 8.4049(5) Å. High-resolution transmission electron microscopy (HRTEM) and its corresponding fast Fourier transform (FFT) were used to identify the crystallographic planes. The crystallographic (113) plane has been found to be one of the most exposed planes in magnetite. Microstructural analysis using both TEM and Field Emission Scanning Electron Microscope (FESEM) demonstrates the formation of well-defined nano-spheres. Energy dispersive X-ray spectroscopy (EDX) confirms the presence of iron (Fe) and oxygen (O) in the sample.

Surface Electronic States

The surface electronic states of distinct elements (Fe, O), as well as surface oxygen vacancies, were investigated using X-ray photoelectron spectroscopy (XPS) using computer program KOIXPD. The background of the core-level spectra was modeled using the Shirley algorithm and the Fe 2p and O 1s core-level spectra were fitted using a least squares non-linear fitting method (Fig. 1, G to I). In this process, the Fe 2p and O 1s core-level spectra were deconvoluted using a combination of Gaussian and Lorentzian profile functions.
The bivalence of iron is evident from Fig. 1, G to H which confirms formation of magnetite. Additionally, the O 1s spectrum reveals a small amount of oxygen vacancies in the sample. The ratio of the area under the curves of $Fe^{2+}$ to $Fe^{3+}$ has been found to be 0.42 after scaling with the photoionization cross-section values specific to the corresponding orbitals. This $Fe^{2+}/Fe^{3+}$ ratio is probably likely due to charge balance on the surface because of the presence of oxygen vacancies.

Magnetoresistance

Hall measurement results yielded a Hall coefficient $R_H$ value of $-4.876 \times 10^7$ cm$^3$/C, indicating the presence of negative charge carriers. Additionally, the magnetoresistance of the magnetite pellet was found to be 3.45 MΩ, which closely aligns with the results observed in our sensing experiments using the magnetite thin-film sensor (~2-4 MΩ). Resistivity of magnetite and mobility of charge carriers have been found to be $6.71 \times 10^5$ Ωcm and 72.7 cm$^2$/Vs respectively.

Magneto-chemiresistance

As illustrated in fig. S10, upon introducing a magnetic field using the bar magnet directed to geomagnetic north, there is an observed increase in the base resistance, indicating a positive total magnetoresistance ($R^D_{TMR}$) of magnetite under the magnetic field along direction D (table S6).

$R^D_{TMR} = R^D_{OMR} + R^D_{AMR} = (R^D_0 - R_0)$ (S3)

Where, $R_0$ = Base resistance when H=0, $R^D_0$ = Base resistance of magnetite for a particular field direction, $R^D_{OMR}$ = Ordinary magnetoresistance with field line along D, $R^D_{AMR}$ = Anisotropic magnetoresistance due to field line along D.

And, then altering the direction of the field line relative to geomagnetic north manifests changes in base resistance in magnetite, indicating the involvement of change in anisotropic magnetoresistance ($R^D_{AMR}$) alongside change in ordinary magnetoresistance ($R^D_{OMR}$). So, these change in base resistance due to the change in field line direction has been denoted as,

$\Delta R^D_{TMR} = (\Delta R^D_{OMR+AMR}) = (R^D_0 - R^N_0)$ (S4)

Where, $R^N_0$ = Base resistance of magnetite when magnet field lines align towards north.

Then equation S5 delineates the concept of total magneto-chemiresistance ($R^D_{TMCR}$), denoting the final resistance of magnetite subsequent to purging it with analyte under a magnetic field along a particular direction. The observed change in resistance, indicating a negative shift in the magneto-chemiresistance of magnetite, is denoted as $\Delta R^D_{MCR}$ (fig. S3 to S6 and S10 and table S6 and S7). It has been observed that alterations in the field strength and field line direction alter the value of $R^D_{TMCR}$ indicating the dependence of magneto-chemiresistance on the strength and direction of the magnetic field (fig. S10 and table S6). This pattern is observed consistently across all directions.

$$R^D_{TMCR} = R_0 + R^D_{TMR} + \Delta R^D_{MCR} \tag{S5}$$

Where, is $R^D_{TMCR}$ = Total magneto-chemiresistance with field lines directed along D, $R^D_{TMR}$ = Total magnetoresistance experienced by the magnetite-based sensor with field line along D, $\Delta R^D_{MCR}$ = Change in magneto-chemiresistance due to field line along D.

$$\Delta R^D_{MCR} = \Delta R^D_{OMCR} + \Delta R^D_{AMCR} + \Delta R_{CR} \tag{S6}$$

Where changes in ordinary and anisotropic magneto-chemiresistance due to field line along D have been denoted as $\Delta R^D_{OMCR}$ and $\Delta R^D_{AMCR}$ respectively and $\Delta R_{CR}$ is the normal chemiresistance change due to analyte.

Changes in magneto-chemiresistance along with changes in magnetoresistance yield distinct magneto-chemiresistive response values along different directions (table S6 to S7).

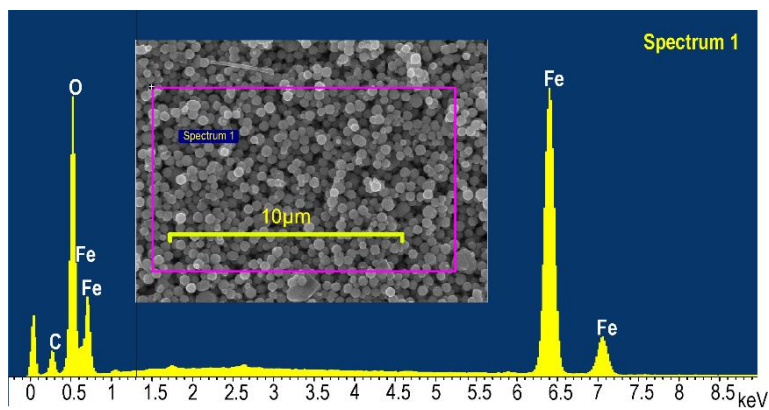

**Fig. S1. EDX spectra of magnetite.**

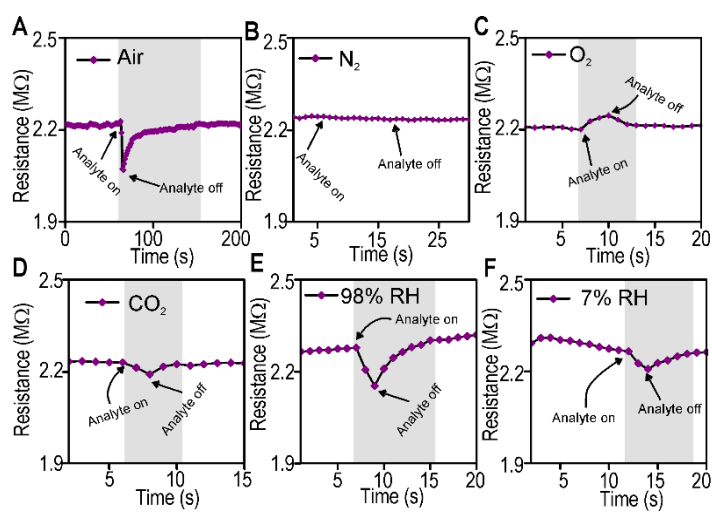

**Fig. S2. Dynamic resistance change curve from chemiresistive sensing experiments using magnetite sensor.** Dynamic resistance changes of magnetite upon purging (**A**) Air. (**B**) Pure nitrogen gas (99.99%). (**C**) 20% oxygen gas (Nitrogen balanced). (**D**) 0.04% carbon dioxide (Nitrogen balanced). (**E**) 98% relative humidity. (**F**) 7% relative humidity.

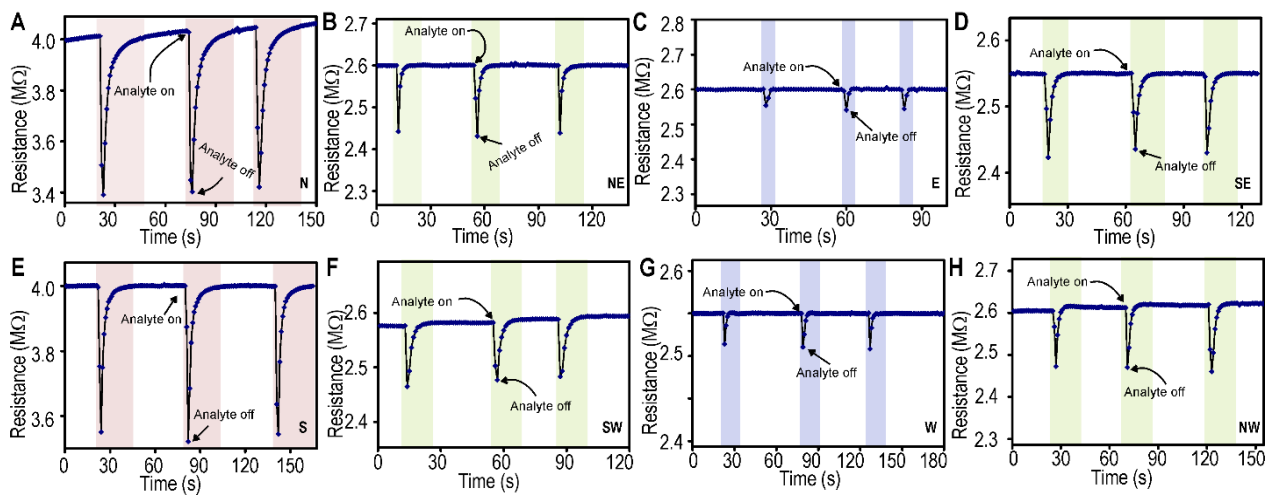

**Fig. S3. Magneto-chemiresistive sensing of magnetite to 98% RH.** (**A**) to (**H**) Dynamic resistance change curve of field line direction dependent magneto-chemiresistive sensing response for different directions (Direction indicated in each fig. are denoted using conventional abbreviations).

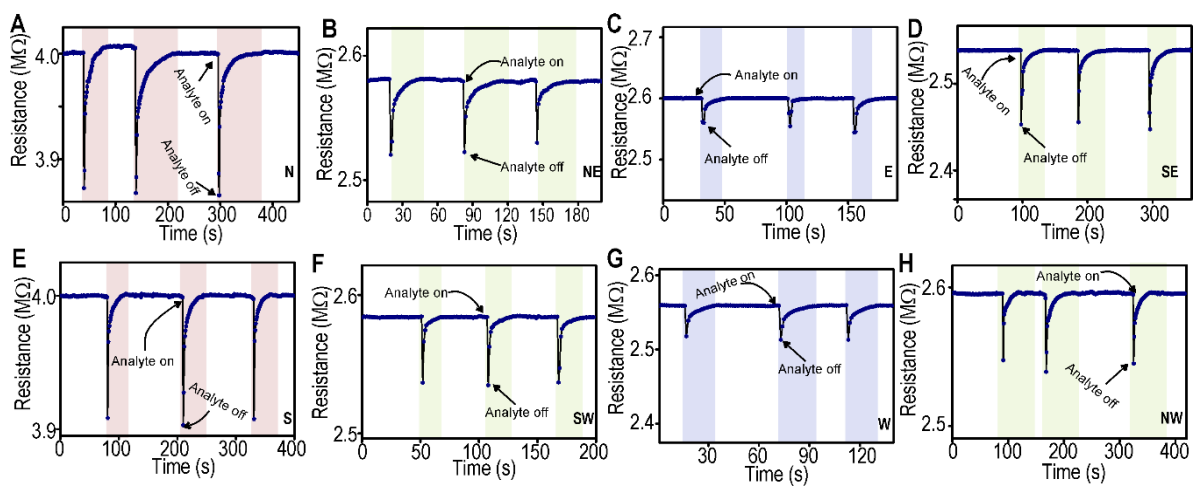

**Fig. S4. Magneto-chemiresistive sensing of magnetite to 7% RH.** (**A**) to (**H**) Dynamic resistance change curve of field line direction dependent magneto-chemiresistive sensing response for different directions (Direction indicated in each fig. are denoted using conventional abbreviations).

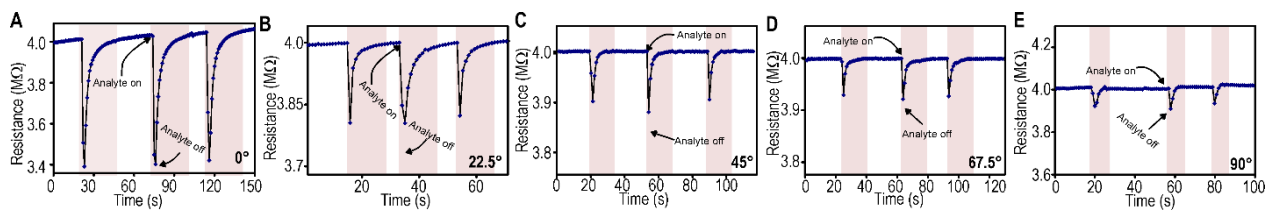

**Fig. S5. Magneto-chemiresistive sensing of magnetite to 98% RH varying inclination angle of magnetite sensor.** (**A**) to (**E**) Plots of dynamic resistance change of magnetite varying inclination angles while bar magnet was directed to geomagnetic north (Inclination angle of sensor is mentioned at right-bottom side of each plot).

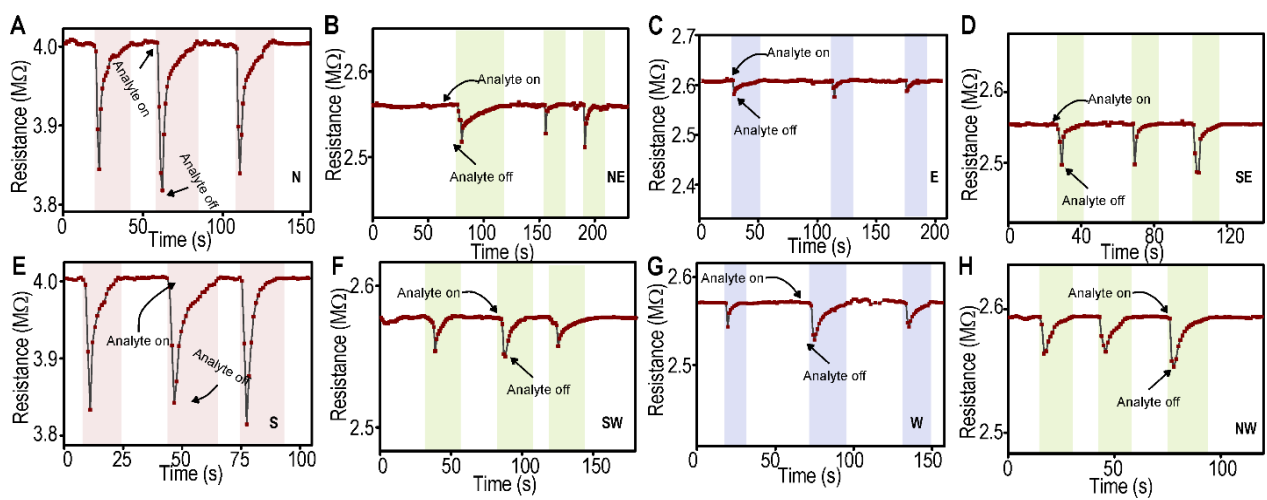

**Fig. S6. Magneto-chemiresistive sensing of magnetite to 98% RH under white light.** (**A**) to (**H**) Dynamic resistance change plots of field line direction dependent magneto-chemiresistive sensing response for different directions under white light (Directions are mentioned at right-bottom sides of each fig. and denoted with conventional abbreviation).

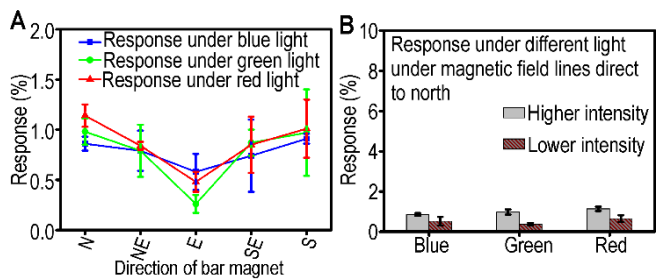

**Fig. S7. Direction dependent magneto-chemiresistive response to 98%RH by magnetite under three primary colors of light.** (A) Changes in magneto-chemiresistive response at different field line direction under blue, green, red light. (B) Changes in magneto-chemiresistive response under blue, green, red light, each with intensities 1800 and 3600 mcd, when field lines of bar magnet were aligned to geomagnetic north.

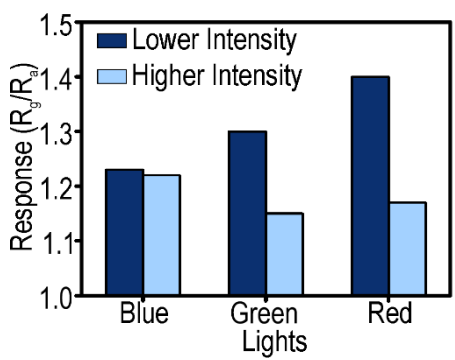

**Fig. S8. Changes in magneto-chemiresistive response to 20 ppm $NO_2$ under blue, green, red light, each with intensities 1800 and 3600 mcd, when field lines of bar magnet were aligned to geomagnetic north.**

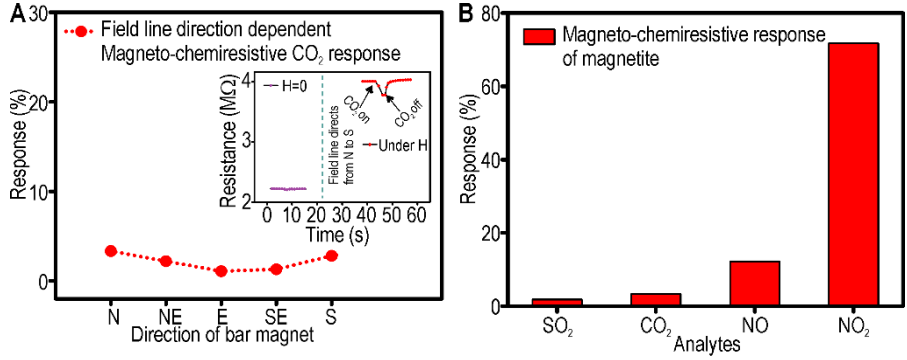

**Fig. S9. Direction dependent magneto-chemiresistive response to CO₂ by magnetite and cross response of magnetite under magnetic field.** Magneto-chemiresistive response for different directions of the bar magnet: (**A**) In presence of CO₂ (inset: dynamic resistance change without magnetic field and under magnetic field where response towards CO₂ has been found 3.33% while the magnetic field lines were directed to geomagnetic north) (**B**) Cross sensing response of magnetite to different toxic analytes where, in presence of NO₂ highest response has been found 71.80% while the magnetic field lines were directed to geomagnetic north.

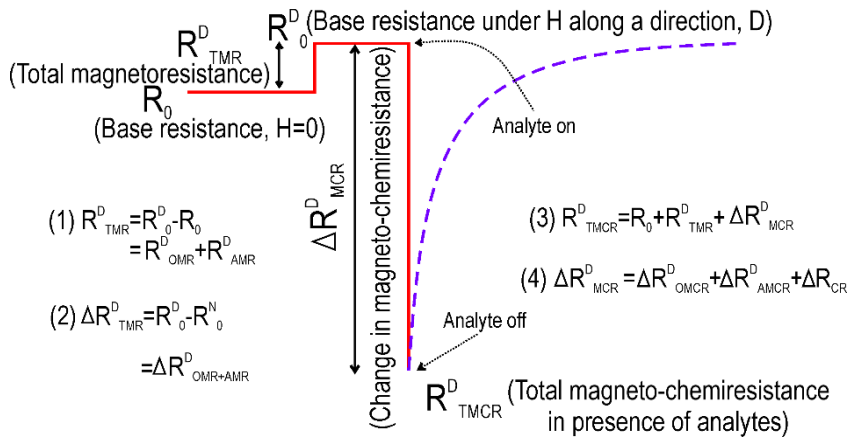

**Fig. S10. Schematic diagram of magneto-chemiresistive resistance and its components.** In magneto-chemiresistive response formula, $R_0$ = Base resistance when H=0, $R^D_0$ = Base resistance of magnetite in a particular direction of field, $R^D_{OMR}$ = Ordinary magnetoresistance with field line along D, $R^D_{AMR}$ = Anisotropic magnetoresistance due to field line along D, $R^N_0$ = Base resistance of magnetite when magnet field lines align towards north, $R^D_{TMCR}$ = Total magneto-chemiresistance in a particular direction, $R^D_{TMR}$ = Total magnetoresistance experienced by the magnetite-based sensor with field line along D, $\Delta R^D_{MCR}$ = Changes in magneto-chemiresistance due to field line along D, $\Delta R^D_{OMCR}$ and $\Delta R^D_{AMCR}$ are changes in ordinary and anisotropic magneto-chemiresistance respectively due to field line along D and $\Delta R_{CR}$ is the normal chemiresistance change due to analyte.

**Table S1: Chemiresistive response of magnetite towards various analytes**

| Analyte | Chemiresistive response value (%/times) |
|---|---|
| Air | 1.70(4) |
| $N_2$ | 0 |
| $CO_2$ | 0.33(5) |
| 98% RH | 0.58(2) |
| 7% RH | 0.48(2) |
| $NO_2$ | 1.028(1) (in times) |
| $O_2$ | 1.013(1) (in times) |

**Table S2: Magneto-chemiresistive response of magnetite towards 98% RH under different magnetic field strength**

| Magnetic Field (mT) | Response (%) |
|---|---|
| 0.02 | 9.40(6) |
| 0.05 | 15.33(26) |
| 1.13 | 31.33(91) |
| 2.6 | 3.74(8) |
| 3.21 | 2.07(8) |

**Table S3: Field line direction and inclination angle dependent magneto-chemiresistive response of magnetite sensor**

| Direction of bar magnet (geomagnetic directions) | Direction dependent Magneto-chemiresistive response (%) | | Inclination angle dependent magneto-chemiresistive sensing | |
|---|---|---|---|---|
| | Without light | Under white light | Inclination angle of sensor (°) | Response (%) |
| North | 15.33(26) | 4.05(39) | 0 | 15.33(26) |
| North East | 5.68(13) | 1.51(22) | 22.5 | 6.65(22) |
| East | 1.25(2) | 0.62(14) | 45 | 2.95(22) |
| South East | 4.95(16) | 1.66(12) | 67.5 | 2.31(8) |
| South | 12.51(25) | 3.96(85) | 90 | 1.86(22) |
| South West | 4.71(9) | 1.16(8) | - | - |
| West | 2.21(27) | 0.72(12) | - | - |
| North West | 4.79(64) | 1.42(16) | - | - |

**Table S4: Magnetic field line direction dependent magneto-chemiresistive response to 7% RH**

| Direction of bar magnet (geomagnetic directions) | Magneto-chemiresistive response (%) to 7% RH |
|---|---|
| North | 2.87(3) |
| North East | 1.70(4) |
| East | 1.20(15) |
| South East | 1.74(12) |
| South | 2.24(5) |
| South West | 1.47(5) |
| West | 1.30(9) |
| North West | 1.64(4) |

**Table S5: Magnetic field line direction dependent magneto-chemiresistive response to 98% RH under different light illuminations**

| Direction of bar magnet (geomagnetic directions) | Response (%) | | |
|---|---|---|---|
| | Blue light | Green light | Red light |
| North | 0.86(7) | 0.98(13) | 1.14(11) |
| North East | 0.79(20) | 0.79(26) | 0.84(4) |
| East | 0.58(18) | 0.26(9) | 0.48(10) |
| South East | 0.74(36) | 0.87(13) | 0.85(28) |
| South | 0.91(5) | 0.97(43) | 1.01(29) |
| West | 0.41(20) | 0.37(6) | 0.51(10) |

**Table S6: Magneto-chemiresistance and its components**

| Direction of bar magnet (geomagnetic directions) | $R^D_{TMCR} = [(R_0 + R^D_{TMR}) + \Delta R^D_{MCR}] = [R^D_0 + \Delta R^D_{MCR}]$ (MΩ) | | | |
|---|---|---|---|---|
| | $R_0$ (MΩ) | $R^D_0$ (MΩ) | $R^D_{TMCR}$ (MΩ) | $\Delta R^D_{MCR}$ (MΩ) |
| Without Magnetic field | 2.228(21) | - | - | - |
| North | - | 4.031(11) | 3.413(19) | -0.618(22) |
| South | - | 4.031(24) | 3.526(23) | -0.504(33) |
| East | - | 2.613(8) | 2.580(8) | -0.033(11) |
| West | - | 2.563(1) | 2.513(6) | -0.050(6) |
| North East | - | 2.569(13) | 2.423(13) | -0.146(19) |
| South West | - | 2.585(1) | 2.464(4) | -0.122(4) |
| North West | - | 2.594(2) | 2.470(2) | -0.124(15) |
| South East | - | 2.532(9) | 2.397(12) | -0.125(15) |

**Table S7: Magneto-chemiresistance and its components**

| Direction of bar magnet (geomagnetic directions) | $\sum_{i=1}^{n}[R^D_{TMR}]_i$ (MΩ) [where, $R^D_{TMCR} = R_0 + R^D_{TMR} + \Delta R^D_{MCR}$] | Chemiresistive detection value of magnetite towards 98% RH $\sum_{i=1}^{n}[\,[R_g-R_a/R_a]_i$ | Magneto-chemiresistive detection value of magnetite towards 98% RH $\sum_{i=1}^{n}[|(\Delta R^D_{MCR})|/R^D_0]_i$ | Change in magneto-resistance (Ordinary + Anisotropic) due to change of bar magnet direction (MΩ) $\Delta R^D_{OMR+AMR}=\sum_{i=1}^{n}[R^D_0-R^N_0]_i$ (MΩ) |
|---|---|---|---|---|
| Without Magnetic field | - | 0.58(2) | - | - |
| North | 1.803(11) | - | 0.1533(26) | 0 |
| South | 1.803(4) | - | 0.1251(25) | -0.001(12) |
| East | 0.385(14) | - | 0.0125(2) | -1.419(3) |
| West | 0.342(20) | - | 0.0221(27) | -1.461(10) |
| North East | 0.341(13) | - | 0.0568(13) | -1.463(5) |
| South West | 0.357(20) | - | 0.0471(9) | -1.446(10) |
| North West | 0.367(20) | - | 0.0479(64) | -1.437(9) |
| South East | 0.304(12) | - | 0.0495(16) | -1.500(3) |